# Vortex-mediated spin current injection into two-dimensional superconductor $NbSe_2$

Meng Yang[1,#], Xiaolong Yin[1,#], Jingjing Liu[1], Yifeng Chen[1], Qinwu Gao[1,2], Hongxing Zhu[1], Danni Huang[1], Lang Chen[1], Shuo-Ying Yang[1,2,*] and Junxue Li[1,2,3,*]

1. Department of Physics, State key laboratory of quantum functional materials, Southern University of Science and Technology, Shenzhen 518055, China
2. Quantum Science Center of Guangdong–Hong Kong–Macao Greater Bay Area (Guangdong), Shenzhen 518045, China
3. Guangdong Provincial Key Laboratory of Advanced Thermoelectric Materials and Device Physics, Southern University of Science and Technology, Shenzhen, 518055, China

# These authors contributed equally: Meng Yang, Xiaolong Yin;

*Corresponding authors: Junxue Li (lijx3@sustech.edu.cn), Shuo-Ying Yang (yangsy@sustech.edu.cn)

**Injection of pure spin current into superconductors remains a major challenge in superconducting spintronics. Previous studies have primarily focused on spin-polarized quasiparticles and spin-triplet supercurrents, while vortices—ubiquitous topological defects in type-II superconductors—have been theoretically proposed as alternative carriers of spin angular momentum, yet direct experimental evidence is still lacking. Here, we report the vortex-mediated spin current injection in $NbSe_2$/$LiAl_2Fe_3O_8$(LAFO) bilayer, an Ising superconductor/ferrimagnetic insulator heterostructure. Under an out-of-plane temperature gradient and an in-plane magnetic field, the $NbSe_2$/LAFO bilayer shows a pronounced thermoelectric peak near the upper critical magnetic field, which has the opposite sign to the conventional vortex Nernst signal observed in a single-layer $NbSe_2$. The sign reversal suggests that the vortex flow induced by spin current injection is opposite to the flow driven by the temperature gradient, which is consistent with theoretical mechanisms including spin-vorticity transmutation and the inverse vortex spin Hall effect. By mapping the field–temperature phase diagram, we reveal that the spin current injection occurs exclusively in the vortex liquid phase of $NbSe_2$. Absence of the thermoelectric signal above the superconducting transition temperature further rules out the quasiparticle contribution. Our results establish vortices as efficient carriers of spin information in superconductors, opening a new route towards vortex-mediated superconducting spintronic devices.**

## Introduction

Realizing long-distance spin transport in superconductors remains a central aim of superconducting spintronics (*1*, *2*). In conventional s- and d-wave superconductors, Cooper pairs form spin singlets (total spin $S = 0$) and therefore cannot carry spin angular momentum. Spin-triplet pairing ($S = 1$) can, in principle, support long-range spin-polarized supercurrents (*3*-*5*), but intrinsic triplet (p-wave) superconductors are exceedingly rare in nature. Alternatively, spin-triplet Cooper pairs can be induced in ordinary superconductors by spin mixing and rotation at magnetically nonuniform interfaces (*6*, *7*), yet producing and unambiguously detecting robust triplet supercurrents remains experimentally difficult. As a result, reports of spin injection into superconductors have generally been attributed to thermally excited, spin-polarized quasiparticles (*8*-*13*).

Vortices are topological defects that emerge in type-II superconductors under an applied magnetic field, consisting of a normal core surrounded by circulating Cooper pairs. Motion of these vortices, driven by electrical currents or thermal gradients, disrupts dissipationless charge transport and produces a finite resistance. Consequently, earlier studies have predominantly focused on enhancing vortex pinning to suppress their motion and thereby increasing the critical current density of superconductors (*14*, *15*).

Recent experiments have demonstrated that superconducting vortex lattices can modulate the spin transport in nearby ferromagnets through magnetic stray fields (*16*-*18*), but the reverse—how injected spin current influences vortex dynamics in superconductors—remains largely unexplored. Theoretical studies suggest that a superconducting vortex can carry angular momentum, either through the orbital circulation of the superfluid around the vortex core (*19*) or via spin polarization of the core induced by the paramagnetic effect (*20*-*22*). In the vortex liquid regime (*23*, *24*), where pinning is weak and vortices move more freely, this motion can facilitate the transfer of spin information. Spin transport by mobile vortices offers two key advantages. First, theory predicts that vortices can exhibit a very large effective spin Hall angle because their drift velocity can greatly exceed the electron drift velocity (*22*), this makes vortices highly efficient converters between spin and charge responses. Second, the topological stability of vortices supports spin transport over relatively long distances (*19*). Together, these properties identify superconducting vortices as promising carriers of spin current.

Experimentally, spin current can be injected from a magnetic insulator into an adjacent nonmagnetic layer by a temperature gradient, as established in longitudinal spin Seebeck experiments (*25*, *26*). In a magnetic insulator/type-II superconductor heterostructure under an out-of-plane temperature gradient, the vortex-mediated spin current (VMSC) signal will be mixed with the thermoelectric response from the conventional vortex Nernst effect (VNE) (*27-29*). However, as illustrated in Fig. 1, the thermoelectric signals from the VNE and vortex-mediated spin-current signals are expected to have opposite sign, so we could qualitatively distinguish these two contributions. Fig. 1A schematically illustrates the VNE in a clean type-II superconductor in the vortex-liquid state. A temperature gradient along z drives vortices from the hot side to the cold side. Phenomenologically (*30*), the thermal force acting on a vortex can be written as $f = -s_{\phi}\nabla T$, where $s_{\phi}$ and $\nabla T$ denote the transport entropy per unit vortex length and the temperature gradient, respectively. In steady state, the balance $\eta \boldsymbol{v}_T = -s_{\phi}\nabla T$ ($\eta$ represents the viscous damping coefficient of vortex motion) gives a vortex velocity $\boldsymbol{v}_T$. Under a magnetic field along y, moving vortices induce an electric field along the x-direction through the Josephson relation (*31-34*), $\boldsymbol{E} = \boldsymbol{B} \times \boldsymbol{v}_T$. In a superconductor/ferromagnetic insulator (SC/FMI) heterostructure (Fig. 1B), spins are injected from the FMI into the superconductor through spin-vorticity transmutation[19] or the spin Seebeck effect (*20-22*), creating a spin-accumulation gradient in the superconducting layer. This gradient acts as a driving force on vortices, producing a vortex flow with velocity $\boldsymbol{v}_S$ and a spin current along z. An electric field along the *x* direction can be established through the Josephson relation $\boldsymbol{E} = \boldsymbol{B} \times \boldsymbol{v}_S$. Notably, the vortex velocity $\boldsymbol{v}_S$ stemmed from spin current injection is opposite to the vortex velocity $\boldsymbol{v}_T$ produced by a temperature gradient, so the electromotive forces from the spin-driven vortex flow and the vortex Nernst effect have opposite signs. Consequently, a pronounced difference in the measured thermoelectric voltage is expected between a bare superconducting layer and an SC/FMI bilayer. However, experimental evidence for vortex-mediated spin injection in SC/FMI heterostructure is still lacking to date.

Here, we report the vortex-mediated spin current injection in an Ising SC/FMI heterostructure, specifically in $NbSe_2$/$LiAl_2Fe_3O_8$(LAFO), where LAFO is a ferrimagnetic insulator. Under an out-of-plane temperature gradient and an in-plane magnetic field, the thermoelectric response of the $NbSe_2$/LAFO bilayer shows a pronounced, sign-reversed peak compared with a pristine $NbSe_2$ layer near the upper critical magnetic field, consistent with vortex flow driven by injected spin current opposing thermally driven vortex motion. The temperature dependence of the thermoelectric signal reveals that both the VNE and vortex-mediated spin current signals are present only in the vortex liquid state and vanish above the critical temperature, allowing us to exclude the quasiparticle contribution. Our experimental observation demonstrates that spin information can be transported in superconductors in the form of vortices, opening new avenues for the development of low-dissipation superconducting spintronic devices.

**Results**

We assemble the SC/FMI heterostructure from van der Waals superconductor $NbSe_2$ and ferrimagnetic insulator $LiAl_2Fe_3O_8$ (LAFO). Conventional high-temperature growth methods for constructing heterostructures often cause interlayer interdiffusion and interface degradation; here, we utilize a two-dimensional (2D) superconductor to effectively avoid this issue, ensuring a cleaner, better-defined interface. $NbSe_2$ is a well-studied 2D superconductor with strong spin-orbit coupling and a range of fascinating properties, including the coexistence of charge density waves and superconductivity (*35*), an in-plane critical field far exceeding the Pauli limit (*36*), an electrically tunable electronic phase transition (*37*), and an orbital Fulde-Ferrell-Larkin-Ovchinnikov (FFLO) state (*38*, *39*). LAFO is a ferrimagnetic insulator with low magnetic damping (*40-42*), making it a promising alternative to yttrium iron garnet (YIG) as a platform for spin-current-related studies. We use a ferrimagnetic insulator instead of a ferromagnetic metal to eliminate spurious contributions from the anomalous Nernst effect (*43-46*) in the magnetic layer during thermoelectric measurements.

High-quality LAFO (001) films, grown by the pulsed laser deposition method, have flat surface and exhibit room-temperature hysteresis (see Supplementary Note 1). Ferromagnetic-resonance measurements give a Gilbert damping $\alpha \approx 3.4\times10^{-3}$ for a 15 nm LAFO film (Supplementary Note 2), consistent with prior work (*40-42*). Longitudinal spin-Seebeck measurements on Pt(3 nm)/LAFO(15 nm) further confirm that these LAFO films are efficient spin-current sources (Supplementary Note 3).

To experimentally investigate vortex-mediated spin current injection, a longitudinal spin Seebeck device with a $NbSe_2$/LAFO heterostructure has been fabricated (Fig. 2A). We pre-patterned Pt electrodes on the LAFO (15 nm) film, then a $NbSe_2$ flake was transferred onto the electrodes and encapsulated with a hexagonal boron nitride (h-BN) layer, to protect the $NbSe_2$ layer from degradation and to electrically isolate it from the top heater. Finally, a Pt heater was defined on top of h-BN by lift-off. For comparison, $NbSe_2$ reference devices were made on nonmagnetic $SiO_2$/Si substrates to measure the conventional vortex Nernst effect (Fig. 2D). Further growth and fabrication details are given in Methods and Supplementary Note 3.

In the thermoelectric measurements (Figs. 2A and 2D), we produce an out-of-plane (*z*-direction) temperature gradient by applying a low-frequency alternating current (AC) to the top heater, in contrast to conventional VNE measurements that use an in-plane gradient. The thermoelectric response along the *x*-direction is detected as the second-harmonic voltage ($V_{2\omega}$) while sweeping an in-plane magnetic field along the $y$-direction. Vortex motion along the $z$-direction induces an electric field along the $x$-direction through the Josephson relation, enabling the direct detection of out-of-plane vortex dynamics.

We first characterize both devices by four-probe resistance measurements in the absence of the heating current. Fig. 2B and 2E show the temperature-dependent resistance of representative $NbSe_2$(19 nm)/LAFO and $NbSe_2$(15 nm)/$SiO_2$/Si samples, respectively, under different magnetic fields. The superconducting transition temperature ($T_c$, defined as the temperature at which the resistance falls to 50% of its normal-state value) is 6.75 K for the $NbSe_2$/LAFO bilayer and 5.54 K for the $NbSe_2$ single layer at zero magnetic field, and shifts to lower temperatures with increasing magnetic field, which agrees with previous reports and can be attributed to the pairing-breaking of the Cooper pairs in a stronger magnetic field. Although interfacial exchange with the ferrimagnetic LAFO(001) film would be expected to suppress superconductivity, here the $NbSe_2$ on LAFO shows a higher $T_c$. One of the possible reasons is that the $NbSe_2$ layer on the LAFO(001) film is slightly thicker; the critical temperature of the $NbSe_2$ flake increases with the layer thickness (*36*).

Fig. 2C and 2F plot the open-circuit thermoelectric voltage ($V_{2\omega}$, cyan color curve) and four-probe resistance ($R$, orange color curve) as a function of an in-plane magnetic field for $NbSe_2$/LAFO and $NbSe_2$/$SiO_2$/Si samples, respectively. To track the superconducting state of the $NbSe_2$ layer during the heating, we measure four-probe resistance simultaneously while driving the top heater. In the $NbSe_2$/LAFO bilayer device (Fig. 2C), $V_{2\omega}$ is negligible at low fields where $NbSe_2$ remains dissipationless, then grows above 5.0 T as $R$ starts to increase, peaks around 7.5 T, and finally decreases to zero. This peak occurs at fields far exceeding the saturation field (~0.15 T) of the 15 nm LAFO film (Supplementary Note 1), excluding the contributions from magnetization reversal and domain-wall dynamics in the ferrimagnetic layer. Instead, the peak aligns with the upper critical field $H_{c2}$ (defined as the field where $R$ reaches 80% of the normal-state value), indicating that it originates from vortex dynamics in the vortex-liquid regime. We define the vortex-lattice-melting-field $\mu_0 H_{\mathrm{melt}}$ as the onset magnetic field at which the thermoelectric voltage becomes finite (as shown by the vertical arrow in Figs. 2C and 2F). The $NbSe_2$/$SiO_2$/Si reference device exhibits a similar peak in $V_{2\omega}$ around the upper critical magnetic field (Fig. 2f), consistent with the conventional VNE. Notably, the thermoelectric voltages in the $NbSe_2$/LAFO bilayer exhibit an opposite sign to that in the $NbSe_2$ single layer, which agrees excellently with the expectation of the vortex-mediated spin injection picture in Fig. 1. This sign reversal is robust and reproduced across multiple $NbSe_2$ devices with and without the ferrimagnetic LAFO (Supplementary Note 5).

Thermoelectric voltages from vortex motion arise only when the superconductor is in a low-dissipation vortex-liquid regime. We therefore map the phase space of the thermoelectric response as a function of magnetic field and temperature. Fig. 3A and 3B show the field dependence of $V_{2\omega}$ at different temperatures for $NbSe_2$/LAFO and $NbSe_2/SiO_2$/Si samples, respectively. The upper critical field $\mu_0 H_{c2}$, determined independently from four-probe resistance measurements under the same heating conditions, is overlaid in Fig. 3A and 3B. For both devices, finite thermoelectric voltage appears only between the vortex-lattice-melting field $\mu_0 H_{\mathrm{melt}}$ and the upper critical field $\mu_0 H_{c2}$. Across the entire temperature range, the $NbSe_2$/LAFO and $NbSe_2/SiO_2$/Si devices display thermoelectric signals of opposite polarity, providing further evidence for the vortex-mediated spin current injection in the $NbSe_2$/LAFO bilayer. With increasing temperature, the voltage peaks in both devices shift toward lower magnetic fields, and the thermoelectric signal vanishes above the superconducting transition temperature, confirming its superconducting origin.

Fig. 3C sketches the temperature–magnetic field phase diagram of a type-II superconductor. Below the superconducting transition temperature, when magnetic fields are below the lower critical field $\mu_0 H_{c1}$, the system resides in the Meissner state and completely expels magnetic flux, vortices are absent, and no vortex-related thermoelectric response is expected. In the intermediate field range between $\mu_0 H_{c1}$ and $\mu_0 H_{c2}$, magnetic flux penetrates the superconductor in the form of quantized vortices, defining the mixed state. In this regime, vortices form a solid phase in which strong pinning suppresses free vortex motion for magnetic fields below the vortex-lattice melting field $\mu_0 H_{\mathrm{melt}}$. Notably, the $\mu_0 H_{c1}$ of the $NbSe_2$ crystal is substantially lower than $\mu_0 H_{\mathrm{melt}}$ where the thermoelectric signals emerge (Supplementary Note 6), ruling out contributions from the vortex-solid phase to the observed thermoelectric signal. The oscillatory thermoelectric voltage observed at low magnetic fields near the superconducting transition temperature in the $NbSe_2/SiO_2$/Si sample (Supplementary Note 5) can be attributed to vortex creeping or hopping (*47*) due to thermal fluctuations or disorder in the vortex solid phase, which agrees with previous report (*48*). However, such mechanisms cannot account for the robust, reproducible sign reversal of the thermoelectric voltage observed in the $NbSe_2$/LAFO heterostructure, where a well-defined polarity is observed over a broad range of magnetic fields. When the magnetic field exceeds $\mu_0 H_{\mathrm{melt}}$, the vortex lattice begins to melt into a vortex liquid, where reduced pinning allows vortices to move freely under thermal gradients or applied currents. This vortex liquid regime is characterized experimentally by a large vortex Nernst signal and a finite resistance. The thermoelectric voltage signals observed in our devices are predominantly located below $\mu_0 H_{c2}$ whereas above $\mu_0 H_{\mathrm{melt}}$, as shown in Fig. 2C and 2F. These features demonstrate that the observed thermoelectric response in our experiments originates from vortex motion in the vortex-liquid state. As a result, we conclude that the spin-driven vortex flow dominates the thermoelectric response in the $NbSe_2$/LAFO device.

In bulk $NbSe_2$ and high-$T_c$ cuprates, quasiparticle can generate a Nernst signal well above $T_c$ with a sign opposite to the VNE signal below $T_c$ (*49*, *50*). To assess possible quasiparticle contributions in our devices, we examine the temperature dependence of the thermoelectric voltage under various magnetic fields for $NbSe_2$ devices with and without the ferrimagnetic LAFO layer (as summarized in Fig. 4A and 4B). The corresponding four-terminal $R$–$T$ curves, obtained under identical heating currents and magnetic fields, are also plotted to identify the superconducting phase. In both $NbSe_2$/LAFO and $NbSe_2$/$SiO_2$ devices, the thermoelectric voltage reaches its maximum magnitude near the midpoint of the superconducting transition, however the two devices exhibit opposite sign to each other over the entire temperature range, consistent with our previous results. Crucially, in our devices the thermoelectric voltage vanishes above the critical temperature, allowing us to unambiguously rule out a quasiparticle origin for the sign reversal of $V_{2\omega}$ in $NbSe_2$/LAFO system. These observations therefore identify vortex-mediated transport as a dominant contribution to the thermoelectric signal in a superconducting layer.

**Discussion**

It is worth comparing the present results with previously reported quasiparticle-mediated spin transport phenomena in superconductors (*11-13*, *51-52*). In those studies, spin angular momentum is carried by nonequilibrium quasiparticles and detected via spin-to-charge conversion. Mobile vortices do not play a role because only small magnetic fields are applied, mainly to switch the magnetization of a ferromagnetic layer adjacent to the superconductor. In contrast, the anomalous thermoelectric response reported here emerges only within the vortex-liquid phase at high magnetic fields, indicating a mechanism that is fundamentally distinct from quasiparticle-mediated spin transport.

Both spin-vorticity transmutation and the inverse vortex spin Hall effect (IVSHE) can, in principle, account for our observations, but they involve different vortex-driving forces. In the spin-vorticity transmutation picture (*19*), the coupling between spin and vorticity is mediated by a normal charge current generated by interfacial spin-orbit interaction, and the spin-driven vortex flow stems from the resulting Lorentz force. In the inverse vortex spin Hall effect picture (*20*, *22*), by contrast, a spin accumulation gradient directly drives vortex motion. In both scenarios, reversing the out-of-plane temperature gradient reverses the direction of spin-driven vortex flow and thus the sign of the thermoelectric signal from vortex-mediated spin-current injection. As demonstrated in the Supplementary Note 7, our control experiments with an inverted temperature gradient are consistent with this expectation and suggest that $V_{VMSC} \propto \nabla T$, whereas the theoretical analysis in Ref. 21 predicts $V_{IVSHE} \propto (\nabla T)^2$. This difference may indicate that the spin accumulation gradient ($\mu_1^{(s)}$) dominates over the spatially uniform component ($\mu_0^{(s)}$) in the regime probed here, but clarifying this point will require further theoretical and experimental work.

Although $NbSe_2$ is an Ising superconductor with strong spin–orbit coupling (SOC) and is therefore expected to support efficient spin–to–charge conversion, we do not observe any inverse spin Hall effect (ISHE) voltage above the superconducting transition temperature in $NbSe_2$/LAFO heterostructures. In the meantime, we find no spin Seebeck signal associated with low-field magnetization switching in LAFO over the entire temperature range, in contrast to evident longitudinal SSE in NbN/YIG heterostructure (*51*) and the pronounced second-harmonic response in Nb nonlocal magnon spin transport device on YIG (*52*). The absence of SSE in our device can be attributed to two possible reasons. First, the $NbSe_2$ layer in our devices is much thicker than the reported spin diffusion length (1 ~ 6 nm) in $NbSe_2$ (*53*), which is typically short in strong-SOC materials. Spin–charge conversion via the inverse spin Hall effect typically increases with the SOC layer thickness, peaks when the thickness is comparable to the spin-diffusion length, and then decreases because of current shunting (*54*). As a result, ISHE signals in our thick $NbSe_2$ flakes are strongly suppressed and become undetectable. Second, thermal-magnon–driven SSE is substantially weaker in heavy-metal/LAFO at low temperatures (Supplementary Note 3), further limiting the detectability of any low-temperature ISHE response. Above all, these considerations highlight that the spin-driven vortex flow provides a more sensitive probe of spin current injection into superconductor than conventional spin–to–charge conversion. By transferring spin angular momentum to vortices, the strong-SOC superconductor utilizes vortices as alternative carriers of spin information, enabling spin transport over length scales that exceed the spin diffusion length. This mechanism opens a promising route to superconducting spintronic devices based on vortex-mediated spin transport.

**Summary**

In summary, we establish a general approach to identify vortex-mediated spin current injection in SC/FMI heterostructures. Using $NbSe_2$/LAFO bilayers, we implement a Nernst-type geometry with an out-of-plane temperature gradient and compare the thermoelectric response with that of a $NbSe_2$ single layer. In the vortex-liquid regime, the $NbSe_2$/LAFO device shows a robust sign reversal of the thermoelectric signal relative to the pristine $NbSe_2$, consistent with theoretical predictions based on spin-vorticity transmutation and inverse vortex spin Hall effect in SC/FMI heterostructures. We would like to add three remarks. First, besides the spin-Seebeck-like method, spin pumping is also an efficient way to inject spin current into type-II superconductor. Second, by Onsager reciprocity, SVT naturally implies its reciprocal effect-a vortex-induced spin torque, calling for further experimental and theoretical studies. Third, given the general nature of vortices in type-II superconductors, the spin-driven vortex flow mechanism is expected to be broadly applicable across a wide range of superconducting systems, potentially extending even to high-$T_c$ superconductors, where vortex physics plays a significant role. Harnessing vortices as carriers of spin angular momentum opens a new route toward long-range spin transport and the development of low-dissipation superconducting spintronic devices.

## Materials and Methods

### Film growth and characterization

$LiAl_2Fe_3O_8$(LAFO) thin films were grown on $MgAl_2O_4$ (001) substrates by pulsed laser deposition (PLD). The base pressure of the growth chamber was better than $9 \times 10^{-9}$ Torr. Prior to loading into the chamber, the substrates were sequentially ultrasonically cleaned in acetone, isopropanol, and deionized water. Before film growth, the substrates were heated in-situ to 550 °C for 20 mins. LAFO films were deposited at an oxygen partial pressure of $1.5 \times 10^{-3}$ mTorr with 5.6% ozone. A 248 nm KrF excimer laser with a pulse energy of 300 mJ and a repetition rate of 1 Hz was used to ablate a stoichiometric LAFO target and generate the plasma plume.

The film thickness was determined by X-ray reflectivity (XRR). Atomic force microscopy (AFM) revealed an atomically flat surface with a root-mean-square roughness below 100 pm over a $5\ \mu m \times 5\ \mu m$ area. Magnetic properties, characterized using a magnetic property measurement system (MPMS), show that the Curie temperature of the 15 nm LAFO film is well above 300 K and the thin film exhibits an in-plane magnetic anisotropy. These properties make LAFO an ideal ferrimagnetic insulator for generating thermally driven spin currents.

### Device fabrication

The bottom electrodes and heater were defined using electron-beam lithography. Poly(methyl methacrylate) (PMMA) resist was spin-coated on the substrates at 4000 rpm and baked at 150 °C for 2 min. For insulating substrates, an additional conductive polymer layer was spin-coated at 2000 rpm to mitigate charging effects during exposure. After electron-beam exposure and development, electrodes were deposited by magnetron sputtering. Contacts to $NbSe_2$ flakes consisted of either 20 nm Pt or 5 nm Ti/10 nm Au (see Supplementary Note 1), whereas the heater electrode was 15 nm Pt.

$NbSe_2$ flakes with thicknesses of 10-20 nm were mechanically exfoliated from bulk crystals (commercially sourced from HQ Graphene) onto $SiO_2$/Si substrates using polydimethylsiloxane (PDMS) films. Hexagonal boron nitride (h-BN) flakes were subsequently exfoliated onto $SiO_2$/Si using the Scotch-tape method. A polycarbonate (PC)-assisted dry-transfer process was then employed to sequentially pick up the h-BN and $NbSe_2$ flakes and place them onto prepatterned electrodes on LAFO or $SiO_2$/Si. All exfoliation and transfer steps were performed in an argon-filled glove box with water and oxygen levels below 0.1 ppm to avoid sample degradation.

**Transport and thermoelectric measurements**

Transport and thermoelectric measurements were carried out using standard lock-in techniques. An alternating current (AC) at 11 Hz from a Keithley 6221 current source was applied to the 15 nm Pt top heater to create a Joule-heating-induced out-of-plane temperature gradient. An in-plane magnetic field, oriented perpendicular to the bottom voltage probes, was swept to drive the $NbSe_2$ layer into the vortex liquid regime. Simultaneously, the second-harmonic voltage in $NbSe_2$ was recorded using SR830 lock-in amplifiers. The heating current ranged from 0.1 to 1 mA. All electric measurements were conducted in a Quantum Design DynaCool system.

**Acknowledgments:** We would like to acknowledge Y.Tserkovnyak at the University of California, Los Angeles for fruitful discussions. **Funding**: M.Y., J.J.L., Y.F.C., Q.W.G. H.X.Z., D.N.H. and J.X.L. were partially supported by National Key R&D Program of China (No. 2025YFA1411302), Key Program of the National Natural Science Foundation of China (NSFC) (No. 12534003), NSFC (No. 12374111), the Shenzhen Science and Technology Program (No. 20220814162144001), Guangdong Provincial Key Laboratory of Advanced Thermoelectric Materials and Device Physics (No.2024B1212010001), Guangdong Major Project of Basic Research (No. 2026B0303000004), and Guangdong Provincial Quantum Science Strategic Initiative (No. GDZX2301004). X.L.Y. and S.Y.Y. were supported by the National Natural Science Foundation of China (No. 12574523, 12404549), Guangdong Fundamental Research Program (No. 2025A1515012039), Guangdong Provincial Quantum Science Strategic Initiative (No. GDZX2402002). L.C. acknowledges the supports by the National Key R&D Program of China (No. 2022YFA1402903), National Natural Science Foundation of China (No. 12574096), Shenzhen Key Laboratory of Phononics and Intelligent Thermal Materials (No. SYSRD20250529114001002), Guangdong Provincial Key Laboratory Program (No. 2021B1212040001) and SUSTech Core Research Facilities. **Author contributions:** J.X.L. and S.Y.Y. conceived and designed the research project. M.Y. deposited LAFO thin films and fabricated the devices with the help of X.L.Y.. M.Y. performed the transport measurements with the help of J.J.L. Y.F.C, Q.W.G.. H.X.Z and D.N.H.. M.Y. analyzed the transport data with the help of X.L.Y., under the supervision of J.X.L. and S.Y.Y. All authors contributed to the writing of the manuscript. **Competing interests**: The authors declare that they have no competing interests. **Data, code, and materials availability**: All data and code needed to evaluate and reproduce the results in the paper are present in the paper and/or the Supplementary Materials. This study did not generate new materials.

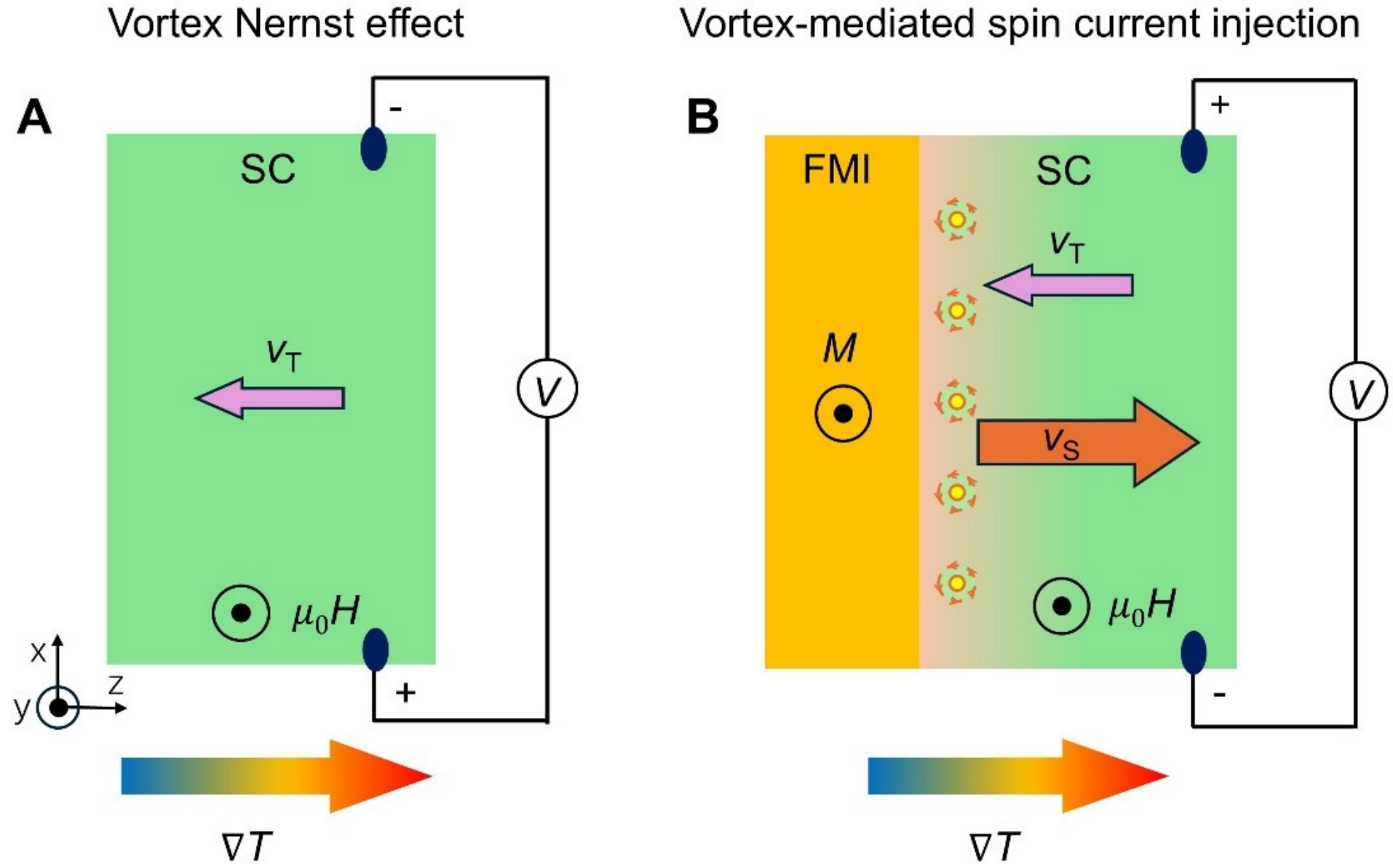

**Fig. 1. Vortex Nernst effect and vortex-mediated spin current injection.** (**A) Vortex Nernst effect**. In a type-II superconductor in the vortex-liquid regime, an out-of-plane temperature gradient ($\nabla T \parallel z$) drives vortices move from the hot to the cold side with velocity $\boldsymbol{v_T}$. In the presence of an in-plane magnetic field ($\mu_0 H \parallel y$), this motion generates a transverse electric field along the *x* direction via the Josephson relation. **(B) Vortex-mediated spin current injection**. In a superconductor/ferromagnetic insulator (SC/FMI) heterostructure, spins are injected from the FMI into the SC by spin-vorticity transmutation or spin Seebeck effect, creating a spin-accumulation gradient in the SC layer that drives vortex motion with velocity $\boldsymbol{v_S}$ along z. Orange circles depict the superconducting vortex at the interface. Because $\boldsymbol{v_S}$ and $\boldsymbol{v_T}$ are opposite to each other, the resulting thermoelectric voltage in the SC/FMI bilayer is expected to have the opposite polarity to the VNE in a single SC layer.

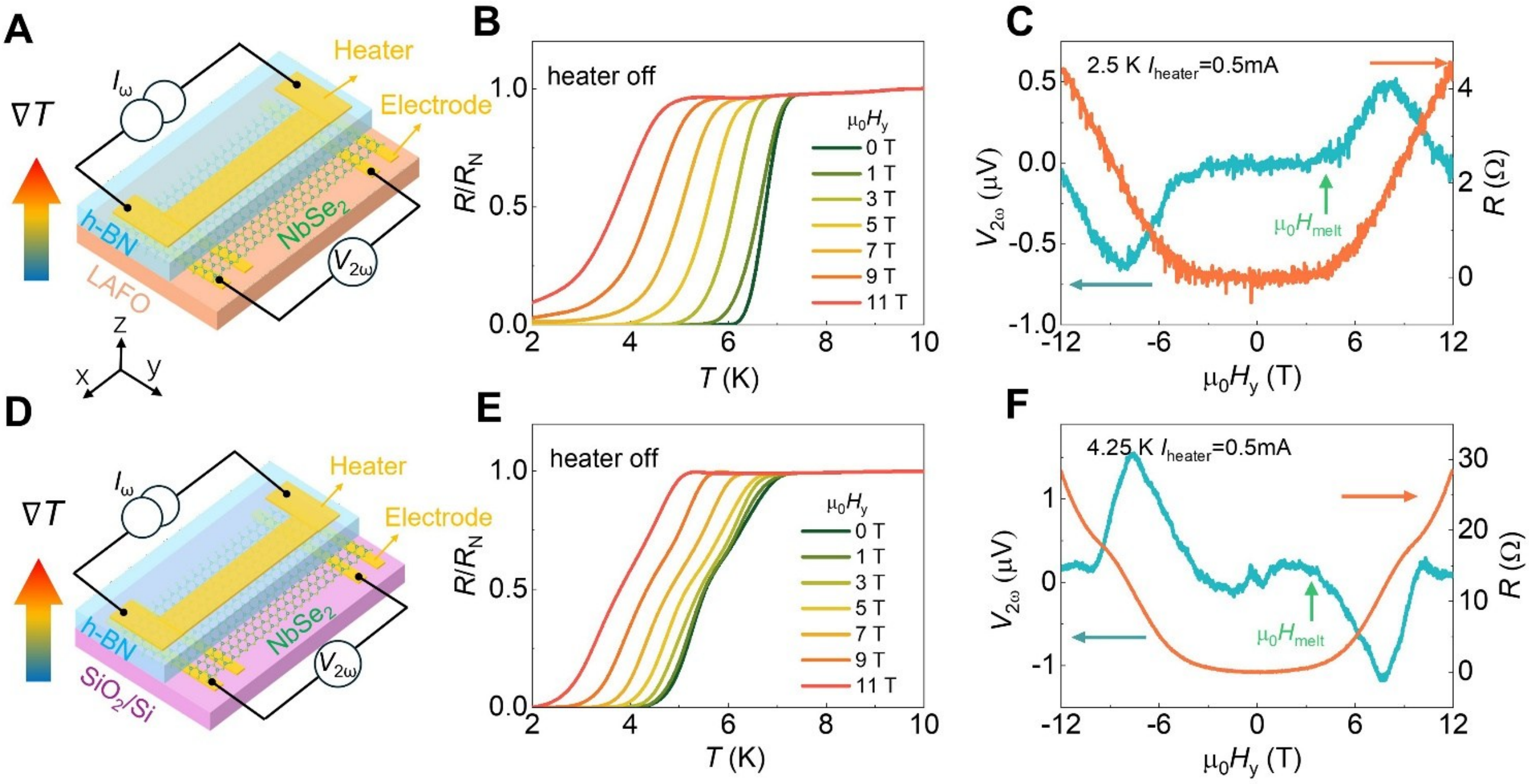

**Fig. 2. Device architecture and thermoelectric transport measurements.** (**A**, **D**) Schematics of the thermoelectric measurement configurations for $NbSe_2$(19 nm)/LAFO(15 nm) and $NbSe_2$(15 nm)/$SiO_2$/Si heterostructures. h-BN/$NbSe_2$ stacks are transferred onto substrates with pre-patterned electrodes. An out-of-plane temperature gradient is generated by an on-chip heater fabricated on top of h-BN, driven by a low-frequency AC current, and the thermoelectric response of $NbSe_2$ is detected as a second harmonic (2ω) voltage. (**B**, **E**) Temperature-dependent four-terminal resistance in the $NbSe_2$/LAFO (B) and $NbSe_2$/$SiO_2$/Si (E) under different magnetic fields applied along the y direction, measured without heating current. (**C**, **F**) Thermoelectric voltage (cyan) and four-terminal resistance (orange) in the $NbSe_2$/LAFO (C) and $NbSe_2$/$SiO_2$/Si (F) as functions of the magnetic field along y-direction under a 0.5 mA heating current. The resistance of the $NbSe_2$ layer is measured with a 1 μA AC probe current while the heater is on. $\mu_0 H_{\mathrm{melt}}$ marks the onset magnetic field at which the thermoelectric voltage becomes finite, indicating the melting of the vortex lattice.

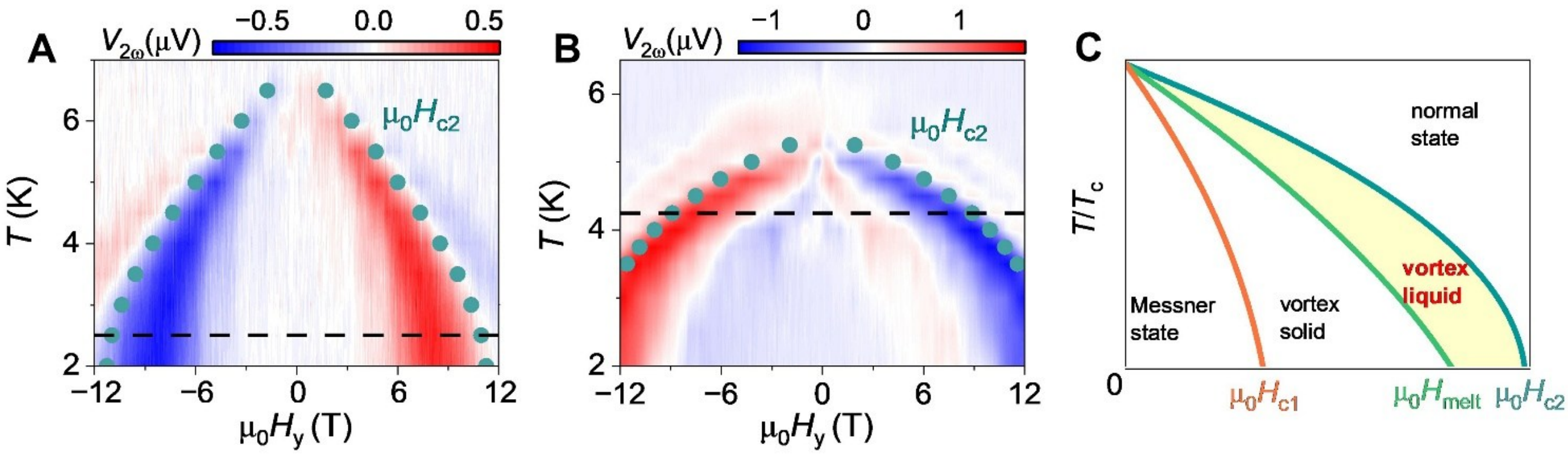

**Fig. 3. Temperature-magnetic field phase diagram of the thermoelectric voltage.** (**A**, **B**) Magnetic field dependence of the thermoelectric voltage for $NbSe_2$(19 nm)/LAFO (15 nm) (A) and $NbSe_2$(15 nm)/$SiO_2$/Si (B) devices at different temperatures, measured with a 0.5 mA current heating. The upper critical magnetic field $\mu_0 H_{c2}$ is extracted from the resistance measurements and defined at 80% of the normal-state resistance. Field-dependent $V_{2\omega}$ at selected temperatures, indicated by dashed lines, are plotted in Fig. 2C and 2F. (**C**) Schematic magnetic field–temperature phase diagram of a type-II superconductor. Here $\mu_0 H_{c1}$ and $\mu_0 H_{c2}$ denote the lower and upper critical magnetic fields, respectively. The vortex solid and vortex liquid phases are separated by the melting field $\mu_0 H_{melt}$.

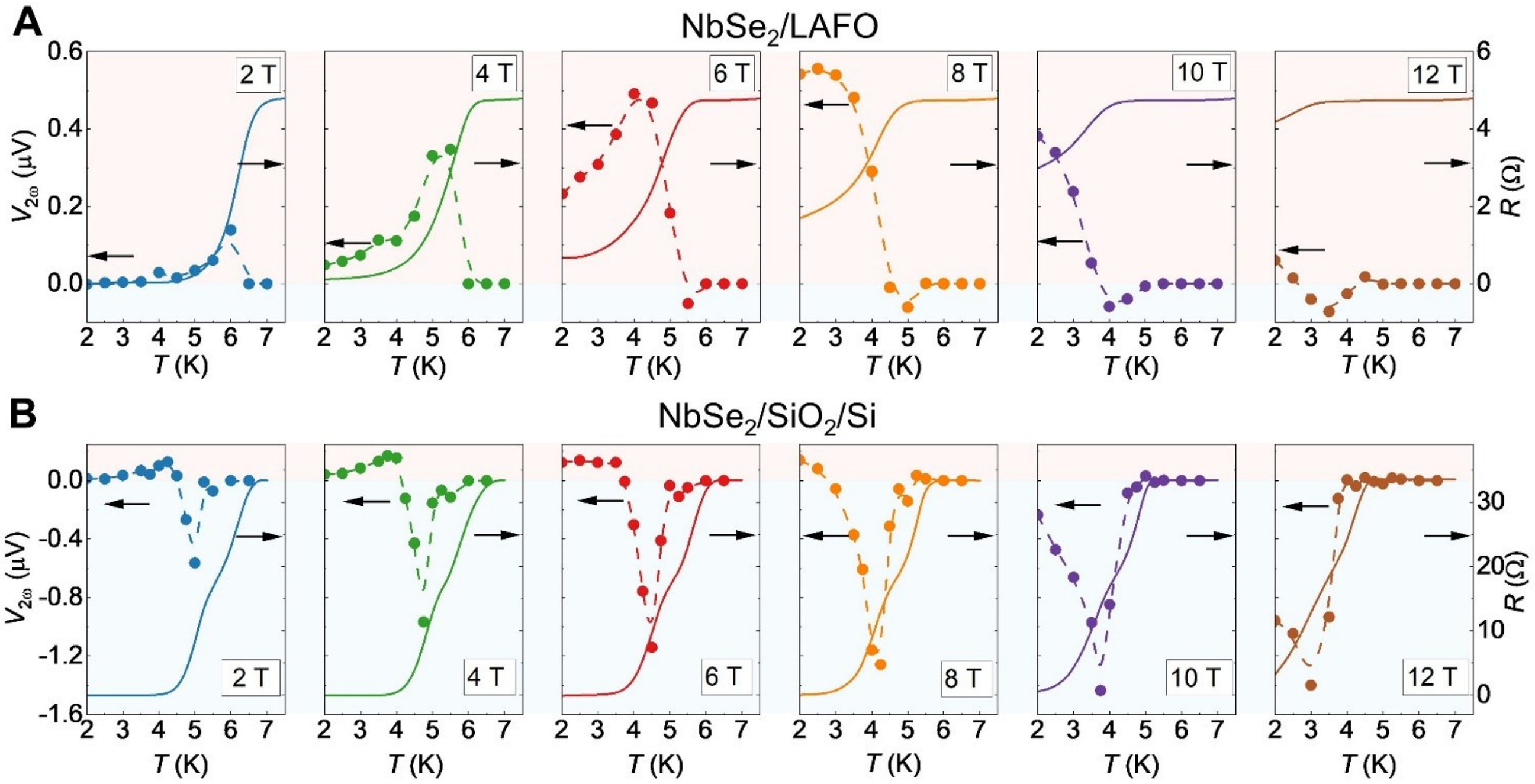

**Fig. 4. Temperature dependence of the thermoelectric voltage.** (**A**, **B**), Thermoelectric voltage as a function of temperature for the $NbSe_2$(19 nm)/LAFO(15 nm) (A) and $NbSe_2$(15 nm)/$SiO_2$/Si (B) devices under selected magnetic fields. The magnitude of thermoelectric signal is extracted by antisymmetrizing the field-dependent- $V_{2\omega}$, $[V_{2\omega}(+H) - V_{2\omega}(-H)]/2$, which removes symmetric thermopower contributions. Dashed lines are guides to the eye; solid curves denote the four-terminal resistance measured under the same heating conditions.